\input{aipcheck}

\documentclass[
    ,final            % use final for the camera ready runs
  ]
  {aipproc}

\layoutstyle{8x11single}
\newcommand {\bc}{\begin{center}}
\newcommand {\ec}{\end{center}}
\newcommand {\bea}{\begin{eqnarray}}
\newcommand {\eea}{\end{eqnarray}}
\newcommand {\be}{\begin{equation}}
\newcommand {\ee}{\end{equation}}

\def\lsim{\mathrel{\rlap{\lower4pt\hbox{\hskip1pt$\sim$}}
    \raise1pt\hbox{$<$}}}               
\def\gsim{\mathrel{\rlap{\lower4pt\hbox{\hskip1pt$\sim$}}
    \raise1pt\hbox{$>$}}}                
\usepackage{graphicx}

\begin{document}

\title{Nearly Perfect Fluidity in the Dilute Fermi Gas: An Update}


\author{Thomas Sch\"afer}{
  address={Department of Physics, North Carolina State University,
           Raleigh, NC 27695}
}

\begin{abstract}
In this contribution we summarize recent results on the 
transport properties of strongly correlated dilute Fermi gases.
We discuss the hydrodynamic equations in the normal phase and 
present new results on the structure of second order terms
in the gradient expansion. We also discuss predictions from 
kinetic theory, and show how these predictions can be tested 
using experimental data on elliptic flow. We summarize current
constraints on the shear viscosity to entropy density ratio 
$\eta/s$.  
\end{abstract}

\maketitle

%%%%%%%%%%%%%%%%%%%%%%%%%%%%%%%%%%%%%%%%%%%%
%% MAINMATTER
%%%%%%%%%%%%%%%%%%%%%%%%%%%%%%%%%%%%%%%%%%%%

%%%%%%%%%%%%%%%%%%%%%%%%%%%%%%%%%%%%%%%%%%%%%%%%%%%%%%%%%%%%%%%%%%%%%%%%
\section{Introduction}
%%%%%%%%%%%%%%%%%%%%%%%%%%%%%%%%%%%%%%%%%%%%%%%%%%%%%%%%%%%%%%%%%%%%%%%%

 In recent years a remarkable convergence has occurred between novel 
physical properties of the hottest matter, the quark gluon plasma, and 
the coldest matter, cold atomic gases, that can be studied in the 
laboratory \cite{Schafer:2009dj}. Both types of matter, which are governed 
by almost scale invariant interactions, were found to behave like nearly 
perfect liquids whose shear viscosity is close to the limits thought to 
be set by quantum mechanics. Initial estimates showed that in both the 
quark gluon plasma and the dilute Fermi gas at unitarity the ratio of 
shear viscosity to entropy density satisfies $\eta/s\lsim 0.5 \hbar/k_B$ 
\cite{Dusling:2007gi,Romatschke:2007mq,Schafer:2007pr,Turlapov:2007}.
This result is close to the value $\eta/s=\hbar/(4\pi k_B)$ which is 
found in the strong coupling limit of a large class of field theories 
that can be analyzed using the AdS/CFT correspondence 
\cite{Policastro:2001yc,Kovtun:2004de}. This suggests that strongly 
coupled quark gluon plasmas and dilute Fermi gases are more efficiently
described in terms of a dual gravitational theory than in terms of a 
conventional quasi-particle model. 

 In this contribution we summarize recent work on transport properties 
of dilute Fermi gases \cite{Bloch:2007,Giorgini:2008}. These systems 
have been realized experimentally using optically trapped alkali atoms 
such as $^6$Li and $^{40}$K. These two atoms are fermions because they 
posses a single valence electron and the nuclear spin is integer.
At very low temperature the atoms can be described as pointlike 
particles interacting via interatomic potentials which depend 
on the hyperfine quantum numbers of the valence electron. A Feshbach 
resonance arises if a molecular bound state in a ``closed'' hyperfine 
channel crosses the threshold of an energetically lower ``open'' 
channel. Because the magnetic moments of the states in the open and 
closed channel are in general different, Feshbach resonances can be 
tuned using an applied magnetic field. At resonance the two-body 
scattering length in the open channel diverges, and the cross section 
$\sigma$ is limited only by unitarity, $\sigma(k) 
= 4\pi/k^2$ where $k$ is the relative momentum. 

%%%%%%%%%%%%%%%%%%%%%%%%%%%%%%%%%%%%%%%%%%%%%%%%%%%%%%%%%%%%%%%%%%%%%%%%
\section{Hydrodynamics}
\label{sec_hydro}
%%%%%%%%%%%%%%%%%%%%%%%%%%%%%%%%%%%%%%%%%%%%%%%%%%%%%%%%%%%%%%%%%%%%%%%%

 At large distances and long times the dynamics of the dilute Fermi 
gas is described by hydrodynamics. For simplicity we will only consider 
the normal phase, corresponding to $T> T_c\simeq0.14 T_F$. Here, $T_F=
k_F^2/(2m)$ is the Fermi temperature, $k_F=(3\pi^2 n)^{2/3}/(2m)$ is 
the Fermi momentum, and $n$ is the density. In the normal phase there 
are five hydrodynamic variables, the mass density $\rho=mn$, the flow 
velocity $\vec{v}$, and the entropy density $s$. These variables 
satisfy five hydrodynamic equations, the continuity  equation, the 
Navier-Stokes equation, and the entropy production equation,
\bea
\label{hydro1}
\frac{\partial \rho}{\partial t} 
   + \vec{\nabla}\cdot\left(\rho\vec{v}\right)  &=& 0 , \\
\label{hydro2}
 \frac{\partial (\rho v_i)}{\partial t}  
   + \nabla_j\Pi_{ij} &=& 0, \\
\label{hydro3}
 \frac{\partial s}{\partial t} 
 \; +\; \nabla_i j_i^{\;s} &=& \frac{R}{T} .  
\eea 
These equations close once we supply constitutive relations for the 
stress tensor $\Pi_{ij}$ and the entropy current $j_i^{\;s}$ 
as well as an equation of state. The constitutive relations can be 
determined order by order in gradients of the hydrodynamic variables.
The stress tensor is given by 
\be 
\label{pi_hyd}
 \Pi_{ij} = \rho v_i v_j + P\delta_{ij}+ \delta \Pi_{ij}\, ,
\ee
where $\delta\Pi_{ij}$, the dissipative part, contains gradients
of $v_i,\rho$ or $s$. At first order the dissipative contribution 
to the stress tensor is $\delta\Pi_{ij}=-\eta \sigma_{ij}$ with
\be 
\label{sig_NS}
 \sigma_{ij} = \left(\nabla_i v_j +\nabla_j v_i 
  -\frac{2}{3}\delta_{ij}(\nabla_k v_k)\right)\, ,
\ee
where $\eta$ is the shear viscosity and we have used the fact that 
the bulk viscosity of the unitary Fermi gas is zero \cite{Son:2005tj}. 
The entropy current is $j_i^{\;s} = v_is+\delta j_i^{\;s}$ with $\delta 
j_i^{\;s}=\kappa q_i$ and $q_i=\nabla_i\log(T)$. Here, $\kappa$ is the
thermal conductivity. Finally, the dissipative function $R$ is given
by $R=(\eta/2)\sigma^2+\kappa q^2$. Scale invariance restricts the 
dependence of the shear viscosity and thermal conductivity on the 
density and the temperature. We can write $\eta(n,T) = \alpha_n(y)n$ 
and $\kappa(n,T) = \sigma_n(y)n/m$, where $\alpha_n(y)$ and $\sigma_n(y)$ 
are functions of $y=mT/n^{2/3}$. 

 We have recently studied the constraints imposed by Galilean invariance 
and conformal symmetry on the structure of second order terms in the 
gradient expansion \cite{Chao:2011cy}. These terms are potentially 
important because the leading order result given in equ.~(\ref{sig_NS}),
which implies that the dissipative stresses are instantaneously 
determined by gradients of the velocity, can lead to acausal behavior
\cite{Schaefer:2009px}. We find that at second order
\bea 
\delta\Pi_{ij} &=& -\eta\sigma_{ij}
   + \eta\tau_R\left(
      \dot\sigma_{ij} + v^k\nabla_k \sigma_{ij}
   + \frac{2}{3} (\nabla^k v_k) \sigma_{ij} \right) 
   + \lambda_1 \sigma_{\langle i}^{\;\;\; k}\sigma^{}_{j\rangle k} 
   + \lambda_2 \sigma_{\langle i}^{\;\;\; k}\Omega^{}_{j\rangle k}
   + \lambda_3 \Omega_{\langle i}^{\;\;\; k}\Omega^{}_{j\rangle k}\nonumber\\  
   && \mbox{}
    + \gamma_1 \nabla_{\langle i}T\nabla_{j\rangle}T
    + \gamma_2 \nabla_{\langle i}P\nabla_{j\rangle}P
    + \gamma_3 \nabla_{\langle i}T\nabla_{j\rangle}P 
    + \gamma_4 \nabla_{\langle i}\nabla_{j\rangle}T 
    + \gamma_5 \nabla_{\langle i}\nabla_{j\rangle}P \, , 
   \label{del_pi_fin}
\eea
where $\tau_R$, $\lambda_i$ and $\gamma_i$ are second order 
transport coefficients, $\Omega_{ij}=\nabla_i v_j -\nabla_j v_i$ is 
the vorticity, and $\langle.\rangle$ denotes the symmetric traceless
part of a second rank tensor. The most important part of this result 
is the structure of the terms proportional to $\tau_R$, which describe 
the relaxation of the dissipative stresses to the Navier-Stokes form. 
Conformal symmetry determines the particular linear combination of 
the comoving derivative of $\sigma_{ij}$ and $(\nabla^k v_k) \sigma_{ij}$ 
that appear in this term.

%%%%%%%%%%%%%%%%%%%%%%%%%%%%%%%%%%%%%%%%%%%%%%%%%%%%%%%%%%%%%%%%%%%%%%%%
\section{Kinetic theory}
\label{sec_kin}
%%%%%%%%%%%%%%%%%%%%%%%%%%%%%%%%%%%%%%%%%%%%%%%%%%%%%%%%%%%%%%%%%%%%%%%%

 Near $T_c$ the transport coefficients $\eta$ and $\kappa$ are 
non-perturbative quantities that have to be extracted from experiment 
or computed in quantum Monte Carlo calculations. At high temperature 
(and at very low temperature, $T\ll T_c$, see \cite{Rupak:2007vp}) 
transport coefficients can be computed in kinetic theory. The shear 
viscosity was first computed in \cite{Bruun:2005}. Here we will follow 
the recent work \cite{Braby:2010xx} which also considers the frequency 
dependence of the shear viscosity. In kinetic theory the stress tensor 
is given by 
\be
\label{T_ij_kin}
\Pi_{ij} = 2\int \frac{d^3p}{(2\pi)^3} 
     \frac{p^i p^j}{m}f_p\, ,
\ee
where $f_p=f(t,x,p)$ is the distribution function of fermion quasi-particles 
and the factor 2 is the spin degeneracy. The equation of motion for $f_p$ 
is the Boltzmann equation. In order to extract the shear viscosity it is 
useful to consider the Boltzmann equation in a curved background metric
$g_{ij}$. In this setting correlation functions of the stress tensor can 
be determined by computing variational derivatives with respect to the 
metric. The non-relativistic limit of the Boltzmann equation in a curved 
space with metric $g_{ij}$ is 
\bea
\label{BE_nr}
\Big( \frac{\partial}{\partial t}
                  + \frac{p^i}{m}\frac{\partial}{\partial x^i} 
  - \Big(  g^{il}\dot{g}_{lj}p^j 
    + \Gamma^{i}_{jk}\frac{p^{j}p^{k}}{m}\Big) 
       \frac{\partial}{\partial p^{i}}\Big) 
    f(t,x,\mathbf{p}) = C[f]\, ,
\eea
where $\Gamma^{i}_{jk}$ is the Christoffel symbol and $C[f]$ is the 
collision integral. We consider small deviations from equilibrium 
and write $f=f_{0}+\delta f$ with $f_{0}(\mathbf{p})=f_{0}(p^{i}p^{j}
g_{ij}/(2mT))$. We also write $g_{ij}=\delta_{ij}+h_{ij}$ and linearize 
in $h_{ij}$ and $\delta f$. We get
\be
\label{boltzmann_linear}
\left(\frac{\partial}{\partial t}
  +\frac{p^i}{m}\frac{\partial}{\partial x^i}\right)\delta f
   +\frac{f_{0}(1-f_{0})}{2mT} p^{i}p^{j}\dot{h}_{ij}
   = C[\delta f]\, . 
\ee
This equation can be solved by going to Fourier space and making 
a suitable ansatz for $\delta f(\omega,k;p)$. Once $\delta f$ is
determined we can compute $\Pi_{ij}$ using equ.~(\ref{T_ij_kin}),
and $\eta(\omega)$ by matching the $k\to 0$ limit to the hydrodynamic 
result equ.~(\ref{pi_hyd},\ref{sig_NS}). The result is  
\be
\label{eta_w}
\eta(\omega) = \frac{\eta(0)}{1+\omega^2\tau^2}\, , 
\hspace{1cm}
\eta(0) = \frac{15 (mT)^{3/2}}{32 \sqrt{\pi}} 
   \left\{
   \begin{array}{cl}
   1 & a\to \infty \\
   1/(3mTa^2) & a \to 0
   \end{array}\right. \,  . 
\ee
We observe that the shear viscosity is large in the weak coupling 
limit $a\to 0$, and that $\eta$ decreases as the scattering length 
is increased. The shear viscosity saturates when the scattering length 
becomes comparable to the de Broglie wave length $a\sim \lambda_{\it dB}
\sim (mT)^{-1/2}$. In this limit $\eta$ only depends on $\lambda_{\it dB}$ 
but not on the density or the scattering length. The frequency dependence 
of the shear viscosity is governed by the relaxation time $\tau = (3\eta)
/(2\varepsilon)$ \cite{Bruun:2007}. We observe that relaxation is fast in 
the limit where the viscosity is small. We also observe that the viscosity 
satisfies a sum rule which only depends on thermodynamic quantities, 
\be
\label{sum_rule}
\frac{1}{\pi}\int_0^{\Lambda} d \omega\, \eta(\omega) 
  = \frac{\varepsilon}{3}\, , 
\ee
where $\Lambda\sim T$ is a cutoff that takes into account the breakdown 
of kinetic theory at very high frequency. The sum rule can be extended 
to $\Lambda\to\infty$ by subtracting the high frequency tail the shear
viscosity, $\eta(\omega)\to\eta(\omega)-C/(15\pi\sqrt{m\omega})$, where 
$C$ is a temperature dependent constant known as the ``contact'' 
\cite{Taylor:2010ju,Enss:2010qh,Goldberger:2011hh,Hofmann:2011qs}.

 There are several other interesting results that have recently 
been obtained in kinetic theory. Chao and Sch\"afer determined 
the second order coefficients defined in equ.~(\ref{del_pi_fin})
\cite{Chao:2011cy}. They find, in particular, that $\tau_R=(3\eta)/
(2\varepsilon)$ agrees with the relaxation time in equ.~(\ref{eta_w}).
Braby et al.~determined the thermal conductivity \cite{Braby:2010ec}, 
and Bruun as well as Sommer et al.~computed the spin diffusion
constant \cite{Bruun:2010,Sommer:2011}. The spin diffusion constant 
at unitarity is 
\be
D_s =\frac{9\pi^{3/2}}{32\sqrt{2}m}
  \left(\frac{T}{T_F}\right)^{3/2}\; . 
\ee
This result can be compared to the momentum diffusion constant 
$D_\eta=\eta/\rho$. Kinetic theory predicts that $D_\eta=45\pi^{3/2}/
(64\sqrt{2}m)(T/T_F)^{3/2}$, which implies that the ratio of $D_s/D_\eta$ 
is independent of temperature. Near $T_c$, both diffusion constants 
are of order $\hbar/m$. A similar relation holds in the quark gluon 
plasma, where the ratio of the heavy quark diffusion constant $D_Q$ 
and the momentum diffusion constant $D_\eta=\eta/(sT)$ is approximately
constant \cite{Adare:2006nq}.

%%%%%%%%%%%%%%%%%%%%%%%%%%%%%%%%%%%%%%%%%%%%%%%%%%%%%%%%%%%%%%%%%%%%%%%%
\section{Elliptic Flow}
%%%%%%%%%%%%%%%%%%%%%%%%%%%%%%%%%%%%%%%%%%%%%%%%%%%%%%%%%%%%%%%%%%%%%%%%

\begin{figure}[t]
\includegraphics[width=.45\textwidth]{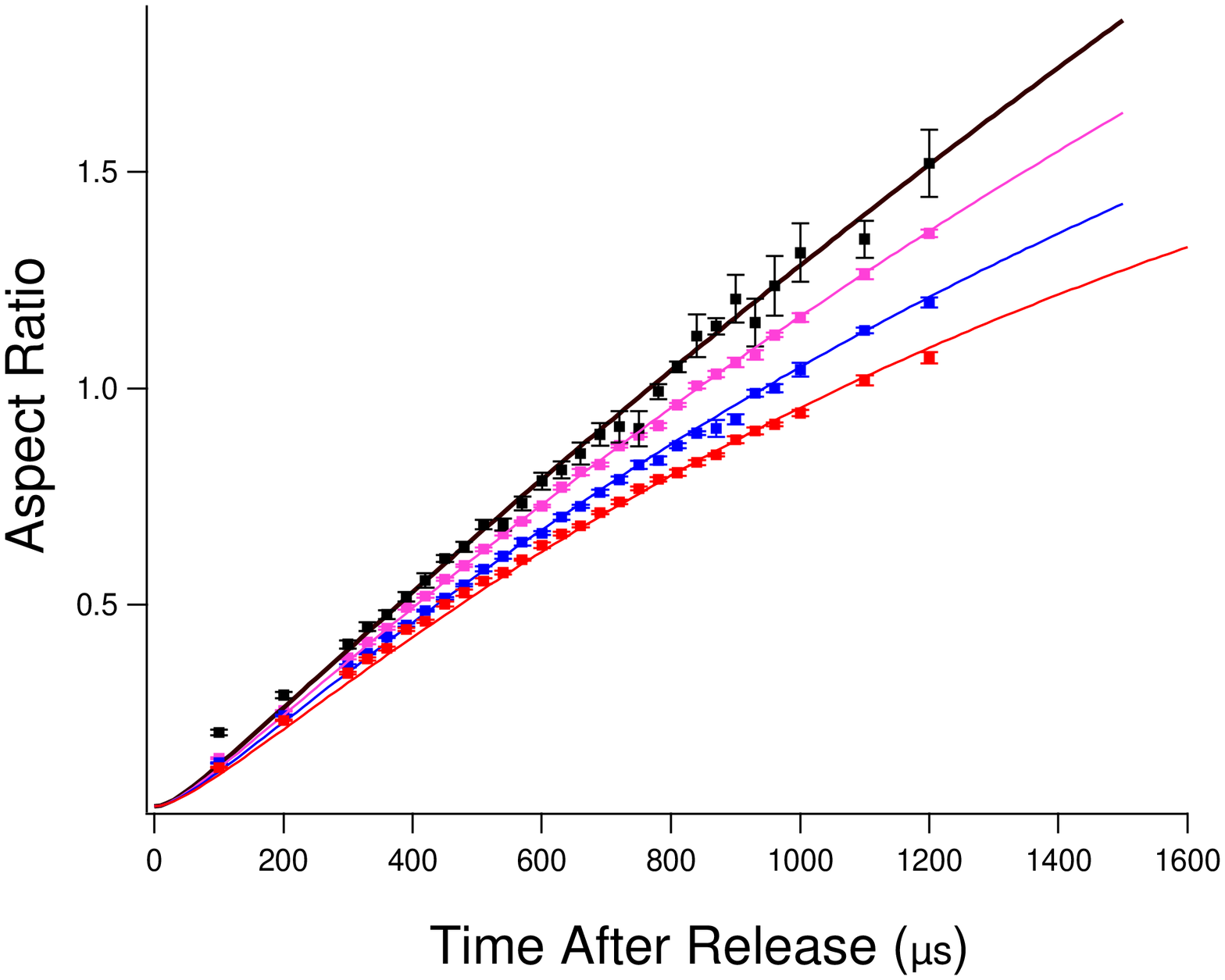}
\includegraphics[width=.45\textwidth]{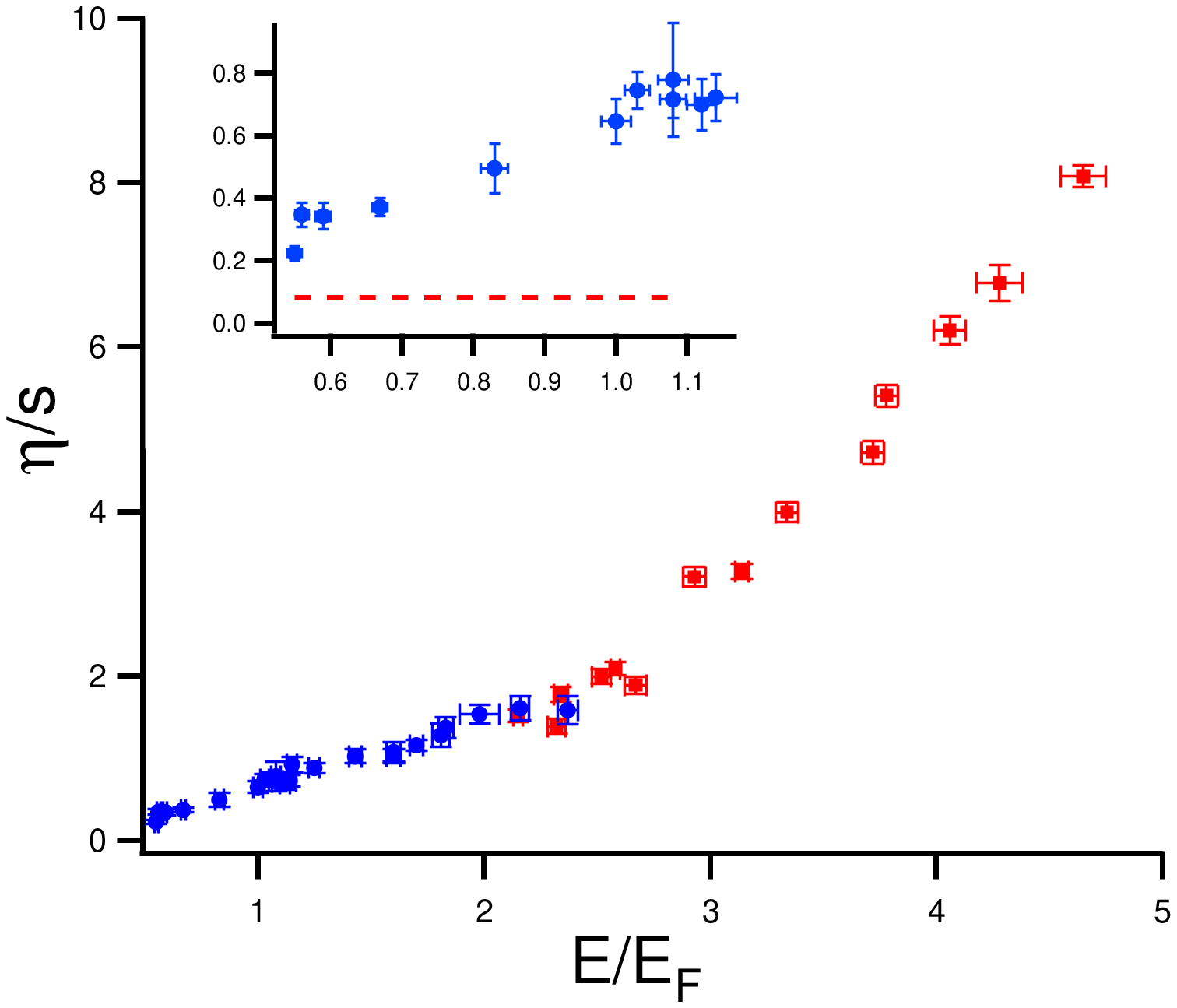}
\caption{\label{fig_A_R_exp}
The left panel shows data for the aspect ratio versus time, from 
\cite{Cao:2010}: Top Black, $E=0.6\,E_F$; Pink,  $E=2.3\,E_F$; Blue, 
$E=3.3\,E_F$ ; Bottom Red, $E=4.6\,E_F$. Solid curves: Hydrodynamic 
theory with the viscosity as the fit parameter. The panel on the 
right shows $\eta/s$ as a function of $E/E_F$. The red data points
come from elliptic flow experiments, the blue data point were
obtained from the damping of collective modes \cite{Cao:2010}. }
\end{figure}

 The first experiment that demonstrated nearly perfect fluidity in 
the dilute Fermi gas was the observation of elliptic flow by 
O'Hara et al.~\cite{OHara:2002}. The experiment involves releasing 
the Fermi gas from a deformed, cylindrically symmetric, trap. The 
density evolves as
\be
\label{n_scale} 
 n(x_\perp,x_z,t) = \frac{1}{b_\perp^2(t)b_z(t)}
  n_0\left(x_\perp b_\perp(t),x_z b_z(t)\right)\, , 
\ee
where $x_\perp,x_z$ are the transverse and longitudinal coordinate, 
$b_\perp(t),b_z(t)$ are scale factors, and $n_0(x_\perp,x_z)$ is 
the equilibrium density of the trapped system. The initial 
system is strongly deformed, $A_R(0)=[\langle x_\perp^2\rangle/
\langle x_z^2\rangle]^{1/2}\ll 1$. Hydrodynamic evolution converts 
the large transverse pressure gradient into transverse flow. As a 
consequence the aspect ratio $A_R(t)$ grows with time and eventually 
becomes larger than one, see Fig.~\ref{fig_A_R_exp}. 

 Viscosity slows down the transverse expansion of the system.
In order to quantify the effect of shear viscosity we have to 
solve the Navier-Stokes equation for the expanding cloud
\cite{Schaefer:2009px,Schafer:2010dv}. In general this has to be
done numerically but for $\eta=\alpha_n n$ with $\alpha_n\simeq
{\it const}$ very accurate semi-analytical scaling solutions
can be found. For this purpose we make a linear ansatz for the 
force term $f_i=(\nabla_i P)/m=a_i x_i$ (no sum over $i$), where 
$a_i=a_i(t)$. The scale parameters $a_i$ and $b_i$ are determined 
by the coupled equations
\bea
\label{ns_for_1} 
\frac{\ddot b_\perp}{b_\perp}  &=&  a_\perp
   -  \frac{2\beta\omega_\perp}{b^2_\perp}
      \left( \frac{\dot b_\perp}{b_\perp} 
                - \frac{\dot b_x}{b_x} \right)\, ,  \\
\label{ns_for_2}
\frac{\ddot b_z}{b_z}  &=& a_z
   +  \frac{4\beta\lambda\omega_z}{b^2_z}
      \left( \frac{\dot b_\perp}{b_\perp} 
                - \frac{\dot b_z}{b_z} \right)\, ,\\
\label{ns_for_3}
\dot{a}_\perp  &=& 
 \!\!\!  -\frac{2}{3}\,a_\perp
   \left(5\,\frac{\dot{b}_\perp}{b_\perp} + \frac{\dot{b}_z}{b_z}\right)
 + \frac{8\beta\omega_\perp^2}{3b_\perp}
  \left(\frac{\dot{b}_\perp}{b_\perp} - \frac{\dot{b}_z}{b_z}\right)^2
  \, , \\
\label{ns_for_4}
 \dot{a}_z  &=& 
  \!\!\! -\frac{2}{3}\,a_z
   \left(4\,\frac{\dot{b}_z}{b_z} + 2\, \frac{\dot{b}_\perp}{b_\perp}\right)
 + \frac{8\beta\lambda\omega_z}{3b_z^2}
  \left(\frac{\dot{b}_\perp}{b_\perp} - \frac{\dot{b}_z}{b_z}\right)^2
  \, ,  
\eea
where $\omega_\perp,\omega_z$ are the oscillator frequencies of the 
harmonic confinement potential (before the gas is released). The 
parameter $\beta$ is defined by 
\be 
\label{beta_def}
\beta = \frac{\alpha_n}{(3N\lambda)^{1/3}}
        \frac{1}{(E_0/E_F)} \, ,
\ee
where $N$ is the number of atoms, $\lambda=A_R(0)$ the initial 
aspect ratio, and $E_0/E_F$ the initial energy in units of 
$E_F=(3N\lambda)^{1/3}N\omega_\perp$. The initial conditions are 
$b_\perp(0)=b_z(0)=1$, $\dot{b}_\perp(0)=\dot{b}_z(0)=0$, and 
$a_\perp(0)=\omega_\perp^2$, $a_z(0)=\omega_z^2$.
 Dissipative effects fall into two categories. The terms proportional
to $\beta$ in equ.~(\ref{ns_for_1},\ref{ns_for_2}) correspond to 
friction -- shear viscosity slows down the expansion in the transverse
direction. The dissipative terms in equ.~(\ref{ns_for_3},\ref{ns_for_4})
describe reheating -- shear viscosity converts some kinetic energy 
to heat which increases the pressure and eventually re-accelerates
the system. Friction slows down the growth of $A_R(t)$ as compared 
to the ideal evolution. Reheating reduces this effect by about a factor 
of 2. 

%%%%%%%%%%%%%%%%%%%%%%%%%%%%%%%%%%%%%%%%%%%%%%%%%%%%%%%%%%%%%%%%%%%%%%%%
%%%%%%%%%%%%%%%%%%%%%%%%%%%%%%%%%%%%%%%%%%%%%%%%%%%%%%%%%%%%%%%%%%%%%%%%

 In order to make comparisons with data we have to take into account
that $\alpha_n$ is not a constant. In \cite{Schaefer:2009px,Schafer:2010dv} 
we argued that $\alpha_n$ in equ.~(\ref{beta_def}) should be interpreted
as the trap average of the local ratio of shear viscosity over 
density, 
\be  
\label{alpha_av}
 \langle \alpha_n\rangle = \frac{1}{N} 
   \int d^3x\, \alpha_n\!\left(\frac{mT}{n_0(x)^{2/3}}\right)n_0(x)\, .
\ee
In the dilute corona of the cloud the local shear viscosity is
independent of the density and equ.~(\ref{alpha_av}) is not well
defined. This problem can be addressed by taking into account that 
the viscous stresses relax to the Navier-Stokes value on a time 
scale $\tau$ that becomes large as the density goes to zero, see
equ.~(\ref{eta_w}). A relaxation model for $\langle\alpha_n\rangle$
was studied in \cite{Schaefer:2009px}. An even simpler model can 
be constructed based on the assumption that the shear viscosity 
relaxes to its equilibrium value at the center of the trap and is 
proportional to the density in the dilute corona. This implies that 
$\eta(x)=\eta(0)(n(x)/n(0))$. This parametrization agrees with the 
relaxation model at the 30\% level. It was used by Cao et 
al.~\cite{Cao:2010} to analyze the data shown in Fig.~\ref{fig_A_R_exp}. 
The hydrodynamic curves shown in Fig.~\ref{fig_A_R_exp} were obtained 
with $\eta=\eta_0 (mT)^{3/2}$ and $\eta_0=0.33$. This agrees quite 
well with the prediction of kinetic theory $\eta_0=15/(32\sqrt{\pi})
\simeq 0.26$, see equ.~(\ref{eta_w}). Once we have verified the 
$T^{3/2}$ scaling behavior of the viscosity in the high temperature 
limit we can determine the shear viscosity to entropy density ratio 
all temperatures, see the right panel in Fig.~\ref{fig_A_R_exp}.
At very low temperature, $T\lsim T_F$, dissipative effects in 
the elliptic flow experiment are very small and it is easier to 
determine the viscosity from the damping of collective modes, see
the blue points in Fig.~\ref{fig_A_R_exp}. We observe that the 
minimum of $\eta/s$ occurs near the lowest temperature studied, and 
that $(\eta/s)_{\it min}\simeq 0.4$. We emphasize that these values
are averaged over the trap, and that more detailed studies are 
needed to find the true minimum of $\eta/s$.

%%%%%%%%%%%%%%%%%%%%%%%%%%%%%%%%%%%%%%%%%%%%%%%%%%%%%%%%%%%%%%%%%%%%%%%%
\section{Conclusions and outlook}
%%%%%%%%%%%%%%%%%%%%%%%%%%%%%%%%%%%%%%%%%%%%%%%%%%%%%%%%%%%%%%%%%%%%%%%%

 There are a number of issues that need to be addressed before an 
accurate value of $\eta/s$ with fully controlled errors can be obtained. 
The most important of these is a better description of the transition 
from nearly perfect fluid dynamics in the center of the cloud to kinetic 
behavior in the dilute corona \cite{Dusling:2011dq}. This problem 
can be addressed using the second order hydrodynamic equations 
discussed in this contribution. We would also like to develop tools 
that will allow us to perform calculations of transport properties in 
the strongly coupled regime. An important constraint for these calculations 
is provided by the sum rules presented in equ.~(\ref{sum_rule}). There 
have also been some very interesting attempts at extending the AdS/CFT 
correspondence to non-relativistic conformally invariant theories, see 
\cite{Balasubramanian:2008dm,Son:2008ye}. Finally, we would like to 
understand whether nearly perfect fluidity in the dilute Fermi gas in 
the strongly interacting regime can be understood in terms of 
quasi-particle degrees of freedom. This question can be addressed
by comparing different transport coefficients, like the ratio
$D_s/D_\eta$ mentioned above, or by computing the spectral functions
related to shear viscosity and diffusion.    

%%%%%%%%%%%%%%%%%%%%%%%%%%%%%%%%%%%%%%%%%%%%%%%%
%% BACKMATTER
%%%%%%%%%%%%%%%%%%%%%%%%%%%%%%%%%%%%%%%%%%%%%%%%

  Acknowledgments: This work was supported in parts by the US Department 
of Energy grant DE-FG02-03ER41260.

\end{document}